\documentstyle[aps,graphicx,floats,prl,amsmath]{revtex}

\begin{document}
\title{Defect Formation and Kinetics of Atomic Terrace Merging} 

\author{ Ajay Gopinathan and T.A. Witten }

\address{Dept. of Physics and James Franck Institute,
The University of Chicago,  Chicago, IL 60637}
\date{\today}

\maketitle

\begin{abstract}
 Pairs of atomic scale terraces on a single crystal metal surface can be made to 
merge
 controllably under suitable conditions to yield steps of double height and 
width.
 We study the effect of various physical parameters on the formation of 
 defects in a kinetic model of step doubling. We treat this manifestly non-
equilibrium problem by mapping the model onto a 1-D random sequential adsorption 
problem and solving this analytically. We also do simulations to check the 
validity of our treatment. We find that our treatment effectively captures the 
dynamic evolution and the final state of the surface morphology. We show that 
the number 
and nature of the defects formed is controlled by a single dimensionless 
parameter $q$.
For $q$ close to one we show that the fraction of defects rises linearly with
 $\epsilon \equiv 1-q$ as $0.284 \times \epsilon$. We also show that one can 
arrive at the final state 
faster and 
with fewer defects by changing the parameter with time.

\end{abstract}

\section{Introduction}
 Structural phase transitions on stepped metals have gained much attention in
 recent years. Starting from some of the first experiments by Lang {\it et al}
 \cite{lang} who noticed single height steps merging to form double height steps 
on Pt\{544\}
 to time lapsed STM measurements of the dynamics of the step doubling process 
more recently \cite{pearl1,pearl2} the field has been attracting more attention. 
The study of
 surface structure and in particular the formation of defects during the step 
doubling process is important as these influence the desirable technological 
properties of the surfaces. Various experimental \cite{pearl1,pearl2,niu,hk} and 
theoretical studies
 \cite{khare,sudoh} have been performed to elucidate the dynamics of the 
evolution of these
 step doubling processes. However one major aspect that has been missing from 
the
 theoretical endeavours is to characterize the evolution of defects in these 
processes. One of our main motivations is to address this question 
theoretically.
\par
 An important characteristic of the step doubling process is that, under the 
optimal conditions, it is manifestly non-equilibrium. Steps seem to double 
irreversibly (unless conditions are changed) and hence equilibrium statistical 
mechanics treatments are suspect. The problem  is thus a part of a whole 
set of interesting problems. These include  non-equilibrium epitaxial growth 
models, many of which are described by the Kardar-Parisi-Zhang equation 
\cite{kardar}. Irreversible depositon of macromolecules on surfaces and 
sociological epidemic models \cite{bailey} also fall in this category. Our 
approach to 
treating the step doubling process is derived from treatments of random 
sequential adsorption (RSA) problems \cite{evans} which deal with irreversible
 adsorption of objects onto lattices.\par
In this paper we consider a simple model \cite{yi} for the step doubling process 
described by two parameters, a nucleation rate and a zippering rate. We first
 map this problem onto a one dimensional random sequential adsorption problem 
with two 
 kinds of species adsorbing onto
 the lattice corresponding to perfectly doubled steps and defective structures.
 We then utilize the methods of RSA to solve the problem analytically. We are 
able to make predictions for the fraction of defects and perfectly doubled steps 
in the asymptotic stages of the process.
We also predict the dynamical evolution of the  degree to which the surface has 
undergone doubling.  The theoretical predictions are then checked against 
extensive simulation results and found to be in good agreement. We finally make 
contact with experimental results and show how our analysis can be used to 
extract information about the dynamics of the process.\par
       The  paper is organized as follows. We first introduce our model for the step doubling process which  utilizes a coarse grained picture of the surface. We then solve the problem in the special limit where the zippering rate is infinite by mapping it exactly to the problem of random sequential adsorption of dimers on a 1-D lattice. We then step back and set up the general mathematical framework for random sequential adsorption of more than one species. We then use this framework to tackle the problem of accounting for defects. We then generalize this to the case where our field of view does not encompass the whole sample but just a part of it, since this is closer to the experimental situation. We then discuss the time dependence of the surface morphology. Next we present our simulation methods and we finally conclude by discussing our results.

\section{Theoretical Analysis}
\begin{figure}
  \centering
  \includegraphics[width=5.0in]{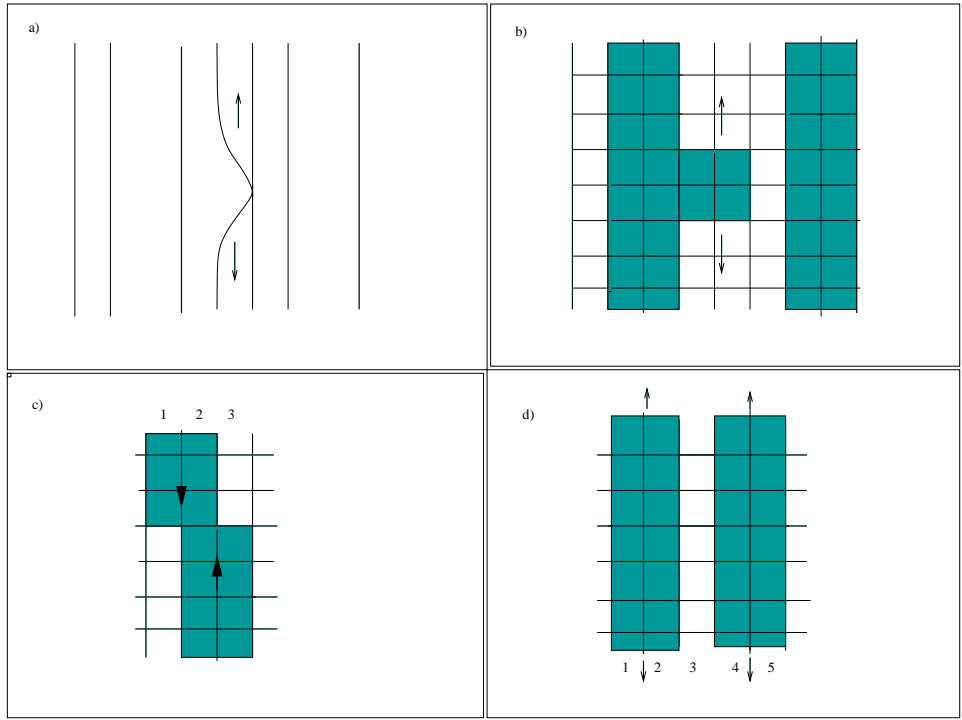}
  \vspace{1ex}
  \caption{(a) A schematic view of single and double steps. The lines represent 
step edges. The thick lines denote double step edges.  The terrace on the left 
of an edge is higher than  the one to the right. Also shown is a single step 
edge bulge touching its downstairs neighbor thus initiating a nucleation. The 
two 
edges will subsequently zip together, as indicated by the arrows, to form a 
double step. (b) A coarse grained lattice representation of the situation in 
(a). Shaded sites represent occupied sites i.e. sites belonging to a double 
step.(c) A frustrated dead end forms when steps 1 and 2 attempt to double and 
steps 2 and 3 also attempt to double starting at another nucleation point. (d) 
An isolated step is formed when steps 1 and 2 and steps 4 and 5 double leaving 
step 3 with no partner to double with. }
  \label{fig:draw}
\end{figure}

  It is found that steps commence coalescing at a point contact 
where a step edge bulge touches the neighboring step edge below 
it\cite{pearl1,pearl2} (see fig.~\ref{fig:draw}). This 
is called a nucleation event. Once a nucleus has been formed the two steps begin 
coalescing steadily in both directions from the nucleus. This process is 
referred to as zippering. Thus there are two distinct rates that govern the 
formation of the doubled structure, the nucleation rate $I$ (defined as the 
number of nucleations that occur per unit time on a sample that has no doubled 
steps) and the zippering rate $Z$ (defined as the  rate of change of the length 
of a doubled step in units of the step width). This zippering rate is observed 
to be large enough that many steps undergo complete doubling. Our analysis will 
focus on this regime of small $I/Z$.
\par
When the step doubling process has reached its final state two main kinds 
of defects are observed to remain. The first kind are isolated step edges whose 
neighbors to both the right and left have doubled with their other neighbors 
leaving the step ``isolated'' (see fig.~\ref{fig:draw}). The second kind are 
frustrated dead ends which are 
formed when a step merges to its left at one point along the step and to its 
right at another point. We wish to understand how the nucleation and 
zippering rates affect the formation of these defects. \par

\begin{figure}
  \centering
  \includegraphics[width=5.0in]{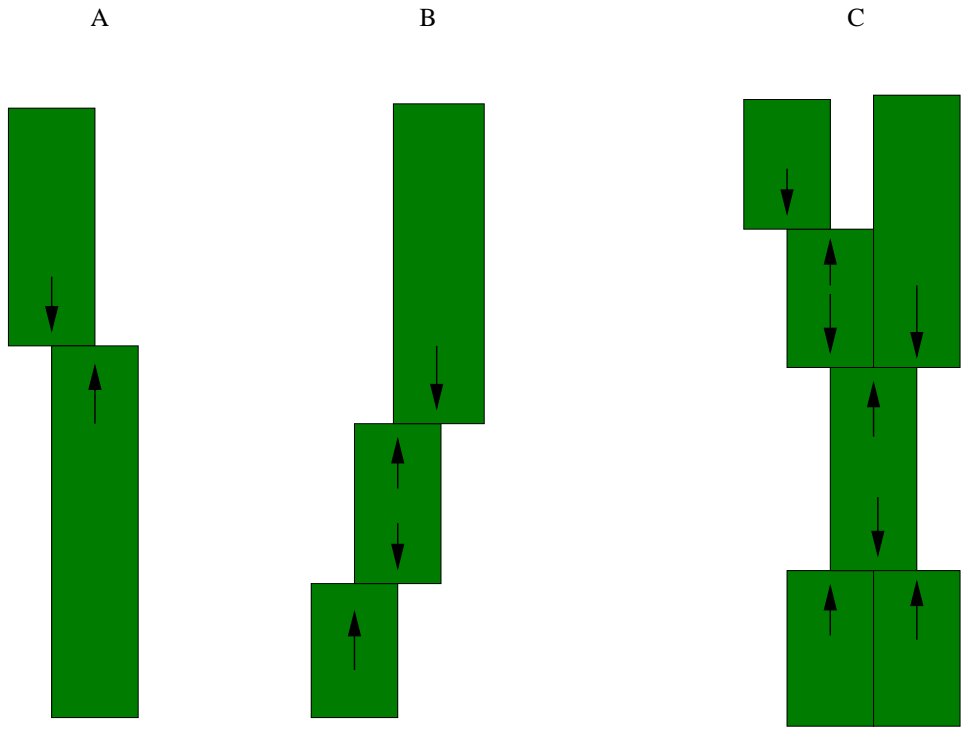}
  \vspace{1ex}
  \caption{Examples of higher order defects: A  defect structure spanning $n+2$ columns is defined to be of order $n$. (A) A frustrated dead end : a defect of order one which spans three columns. (B) A defect of order two spanning four columns.(C) A defect of order three spanning five columns.}
  \label{fig:defect}
\end{figure}
\subsection{Final State}

 The simplest model assumes that nucleation points occur {\it randomly} over the 
entire sample with a probability per unit time determined by the nucleation rate 
($I$). The steps then coalesce at a steady rate ($Z$). To attack this problem we 
first imagine mapping a square section of the sample to a coarse grained lattice 
(see fig.~\ref{fig:draw}).
Here each column represents a step. A lattice site can either be occupied or 
unoccupied. Occupied sites represent sections of a step that have doubled. A 
step is allowed to double only with its downstairs neighbor (to the right as in 
fig.~\ref{fig:draw}). Thus occupied lattice sites occur in pairs, one each on 
adjacent steps. A nucleation event where a step edge bulge touches its 
downstairs neighbor is represented as shown in fig.~\ref{fig:draw} in the 
coarse grained picture. Taking the coarse grained view involves the loss of some 
specific information about the system. One loses information about the exact 
position of a nucleation point and also the exact shape of the doubling step. 
Positions in the vertical and horizontal direction can only be specified up to a 
single step width. Thus the coarse graining will also fail to capture the 
meandering of individual steps. \par
 Once a 
nucleus has been formed the two steps will eventually form a double step {\it 
unless} it gets frustrated as described previously (and in fig.~\ref{fig:draw}). If we examine the system after all evolution has ceased, there 
are thus at least two 
possible entities: a perfectly doubled step which occupies two columns of the 
coarse grained lattice and a 
frustrated pair which occupies three columns. There are in principle higher 
order defects that can occupy more columns (see fig.~\ref{fig:defect}). For example one can have an extended 
structure where the doubling of steps 1 and 2 is frustrated by the doubling of 
steps 2 and 3 at another point which in turn is frustrated by steps 3 and 4 
nucleating a doubling event and so on. To be able to proceed with 
an analysis we need to understand what kind of defects are important and under 
what conditions. We first consider the simplest case where the zippering rate is
taken to be infinite.

\subsection{Infinite Zippering Rate}
In this case the nucleation rate, $I$, is assumed to be some finite number while 
 the zippering rate, $Z$, is taken to be infinite. Now {\it any} nucleation will 
lead to a perfectly doubled step as the steps zipper together instantaneously.
 There will be no frustrated dead ends in this scenario. However there will be 
isolated steps, both of whose neighbors have doubled with other steps. The
question then is how many of these isolated  steps will be present in the final 
state when all evolution is complete.\par
To answer this question, we first note that a nucleation attempts the occupation 
of two columns in the coarse grained lattice picture. If the columns are 
unoccupied before the attempt, the nucleation is successful and both columns are 
occupied.  Thus in any given time step when a succesful nucleation occurs two 
adjacent columns are fully occupied. Furthermore
once these columns have been occupied subsequent successful nucleations cannot 
overlap either of these columns. We can now visualize the process as occuring on 
a 1D lattice
each site of which corresponds to a column on our original lattice.  A 
successful nucleation event leads to the occupation of two consecutive sites on 
this 1D lattice.  Each nucleation attempt therefore corresponds to a random 
choice of a pair of consecutive sites. If both these sites are unoccupied, the 
attempt is successful and as a result both the sites are occupied.  This process 
is {\it exactly} random sequential adsorption (RSA) of dimers (objects that 
occupy two lattice sites) onto a 1D lattice. In the RSA process a dimer 
attempts to adsorb onto the lattice at every time step, corresponding to the 
nucleation attempts in our system. Dimers cannot overlap just as in our system 
already doubled steps cannot double again with a different step. Thus in the 
limit of an infinite zippering rate our problem maps {\it exactly} to an RSA 
problem of dimers adsorbing onto a 1D lattice.\par
Though the solution to the RSA problem is standard \cite{evans,flory}, we 
present the solution for completeness and to set the stage for the next 
subsection. We first define $P_{n}$ to 
be the probability that a randomly chosen site is part of a sequence of {\it at 
least}
 $n$ consecutive empty 
sites.  We can then write down the time evolution of this probability for $n 
\geq 
1$
\begin{equation} 
  \frac{dP_{n}}{dt} = -k (n-1) P_{n} - 2 k P_{n+1} 
\label{eq:master0}
\end{equation}
The first term accounts for the sequence being broken by the adsorption of a 
dimer 
within the $n$ sites, which can be done in $n-1$ ways. The second term refers to 
a 
dimer overlapping the sequence from either end if the sequence has at least 
$n+1$ sites.
This can be done in two ways. In order to solve this set of equations, we first define the conditional probability $g_n = P_n/P_{n-1}$. Using equation \ref{eq:master0}, we can write down the set of equations satisfied by the $g_n$'s.
\begin{equation} 
\frac{dg_{n}}{dt} = -k g_{n} - 2 k (g_n g_{n+1} - g_n^2)
\label{eq:masterg0}
\end{equation}
Now at time $t=0$, the lattice is empty and all the $P_n$'s are identically unity. This implies that all the $g_n$'s are unity at $t=0$. Therefore the initial conditions tell us that all the $g_n$'s satisfy the same equation at $t=0$ since the second term on the right hand side in equation \ref{eq:masterg0} does not contribute. However this means that as time progresses the $g_n$'s evolve in an identical fashion, all of them being equal, with the second term {\it never} contributing. Thus $g_n = f(t)$ is simply a function of time and does not depend on $n$. We can therefore look for solutions to equation \ref{eq:master0} of the form
\begin{equation} 
P_{n} = f(t) P_{n-1}
\end{equation}
 Using this with eq.(\ref{eq:master0}) gives a differential equation for $f(t)$ (which can also obtained directly from eq.\ref{eq:masterg0})
that may be solved to yield the {\it ansatz}
\begin{equation} 
P_{n} = e^{- k (n-1) t} P_{1}
\label{eq:ansatz0}
\end{equation}
It may be readily verified that the {\it ansatz} consistently satisfies 
eq.(\ref{eq:master0}). Now using (\ref{eq:master0}) and 
(\ref{eq:ansatz0}) we get 
\begin{equation}
\frac{dP_{1}}{dt} = - 2 k e^{-kt} P_{1}
\end{equation}
which yields
 \begin{equation}
  \ln P_{1} =  2 e^{-(k)t} + c
\end{equation}
 where $c$ is an arbitrary constant. Using the initial condition that at $t=0$, 
the lattice is empty and hence $P_{1}=1$ we get
 \begin{equation}
 c = -2
 \end{equation}
We thus have an explicit solution for $P_{1}$ 
 \begin{equation}
   P_{1} =  \exp [ 2 e^{-(k)t} - 2]
 \label{eq:infsol}
\end{equation}
This tells us that when $t \rightarrow \infty$, $P_{1} \rightarrow e^{-2}$. The 
fraction of sites that are unoccupied in the final state is therefore $e^{-2} 
\approx 13.5 \%$. Thus even in the case where we only have perfectly doubled 
steps the percentage of total area covered by the doubled steps is $86.5 
\%$. There will {\it always} be $13.5 \%$ of the surface covered by isolated 
single step defects. In the case of a finite system we expect $13.5\%$ defective area on average. We will later consider the case where the ratio $I/Z$ is 
small but finite. Here we expect not only isolated step edges but also 
frustrated dead ends and possibly defects of higher order.

\subsection{Accounting for frustrated dead ends }

 In real experimental situations \cite{pearl1,pearl2,yi}, one observes not only perfectly doubled steps but also frustrated dead ends (defects of order one). In this section we will lay down the mathematical framework, which we will use to treat the occurence of defects. {\it Suppose} we are in a regime (low $I/Z$) where we have {\it only} two significant entities: perfectly doubled steps and frustrated dead ends. A perfectly doubled step occupies two columns and a frustrated dead end spans three columns. We may now think of the process as occuring on a 1-D lattice with dimers (blocks of two lattice units) representing perfectly doubled steps and trimers (blocks of three lattice units) representing frustrated dead ends adsorbing onto the lattice at different rates. The dimer attempt rate is denoted by 
$k$ and the trimer attempt rate by  $k'$. These attempt rates are in general functions of time, the nucleation rate and zippering rate. The problem of determining these attempt rates will be dealt with in the next section. For now we take them to be arbitrary functions of time. The kinetics of this process corresponds to the problem of random sequential adsorption 
(RSA) of a binary mixture \cite{bonnier,bartelt} though with attempt rates that are functions of time. As before we first define $P_{n}$
 to 
be the probability that a randomly chosen site is part of a sequence of {\it at 
least}
 $n$ consecutive empty 
sites.  We can then write down the time evolution of this probability for $n > 
2$
\begin{equation} 
  \frac{dP_{n}}{dt} = -k (n-1) P_{n} - 2 k P_{n+1} -k' (n-2) P_{n} -2 k' P_{n+1} 
- 2 k' P_{n+2}
\label{eq:mastert}
\end{equation}
The first term accounts for the sequence being broken by the adsorption of a 
dimer 
within the $n$ sites, which can be done in $n-1$ ways. The second term refers to 
a 
dimer overlapping the sequence from either end if the sequence has at least 
$n+1$ sites.
This can be done in two ways. The third, fourth and fifth terms are similar to 
the 
first and  second except that they refer to the adsorption of trimers in an 
analogous
fashion. Introducing the conditional probabilities $g_n = P_n/P_{n-1}$  and using the initial conditions, we find that $g_n=f(t)$ is only a function of time and does not depend on $n$.
We therefore introduce the ansatz 
\begin{equation} 
P_{n} = f(t) P_{n-1}
\label{eq:ansatzt}
\end{equation}
Using this in conjunction with equation (\ref{eq:mastert}) and the initial condition 
$P_n=1$ yields
\begin{equation}
  f(t) = e^{-\int_{0}^{t} k(t')+k'(t') dt'}
\end{equation}
This along with equations (\ref{eq:mastert}) and (\ref{eq:ansatzt}) yield an 
explicit solution for $P_2$
\begin{equation}
  P_2 = \exp \left[-\int_{0}^{t} k(t') dt' - \int_{0}^{t}[ 2 f(t') (k(t')+k'(t'))+2k'(t')(f(t))^2] dt'\right]
 \label{eq:P2t}
\end{equation}
and hence from (\ref{eq:ansatzt}) for all $P_n$ for $n \ge 2$. For $n=1$  we
have
 \begin{equation}
  \frac{dP_{1}}{dt} = -2kP_{2} - 3k'P_{3}
  \end{equation}
which upon integration gives
 \begin{equation}
  1 - P_{1}(t=\infty) = \int_{0}^{\infty}( 2kP_{2} +3k'P_{3} ) dt
 \label{eq:asymfract}
 \end{equation}
$P_{1}(t=\infty)$ is simply the fraction of space occupied by isolated dead ends 
at asymptotic coverage. The two terms on the right are simply the fraction of 
space occupied by the dimers (double steps) and trimers (frustrated dead 
ends) respectively.
It is to be noted that this constitutes an exact solution for the problem of random sequential adsorption of a binary mixture in 1D with {\it arbitrary} time dependence of the attempt rates. It is also to be noted that this analysis may be extended to include defects of higher order if one knows the attempt rates for all the species being considered. We now consider the case with $I/Z$ small but non-zero and compute the rates $k$ and $k'$ explicitly in terms of the experimental parameters $I$ and $Z$.

\subsection{Nonzero but Small I/Z}
We now apply the formalism developed in the previous section to our problem where we are given the experimental parameters: the nucleation  rate $I$ and the zippering rate $Z$.
The case when $I/Z$ is small but non-zero is important as it is closer to 
reality. Most experimental situations have $I/Z$ values which are typically of 
order $ 10^{-3}$ to $10^{-1}$ \cite{sudoh,yi}. To begin our analysis we need to 
understand for what values of $I/Z$ different orders of defects become 
important. Intuitively one expects that for small enough $I/Z$, considering only 
defects of order one would be a good approximation. To estimate how small $I/Z$ 
needs to be we first consider the mean free path, $ \langle l \rangle$, of a successful nucleation on 
an empty lattice.  By mean free path we mean the average length to which the 
doubled step grows before it is cut off by other nucleations.

Heuristically one can say that higher order defects will not be important when 
the mean free path is much larger than the size of the lattice we are dealing 
with i.e. $ \langle l \rangle \gg M $. We now focus on a
regime of low $I/Z$ that satisfies the above condition with the assumption that we only have frustrated dead ends. The explicit condition will be worked out later in this section.
\par
We can now view the whole process as occuring on a 1-D lattice of size $M$ corresponding to the problem of random sequential adsorption 
(RSA) of a binary mixture.
In our system there are only dimers trying to 
adsorb onto the lattice except they may turn into a trimer if another dimer 
overlaps one of its units within the time it takes to zipper $M$ steps. We thus 
view this as a RSA process with the rate at which trimers try to adsorb 
 determined by the probability, $q$, of the above mentioned overlap occuring. 
We now 
compute this probability explicitly in terms of $I$ and $Z$.
\par
\begin{figure}
  \centering
  \includegraphics[width=5.0in]{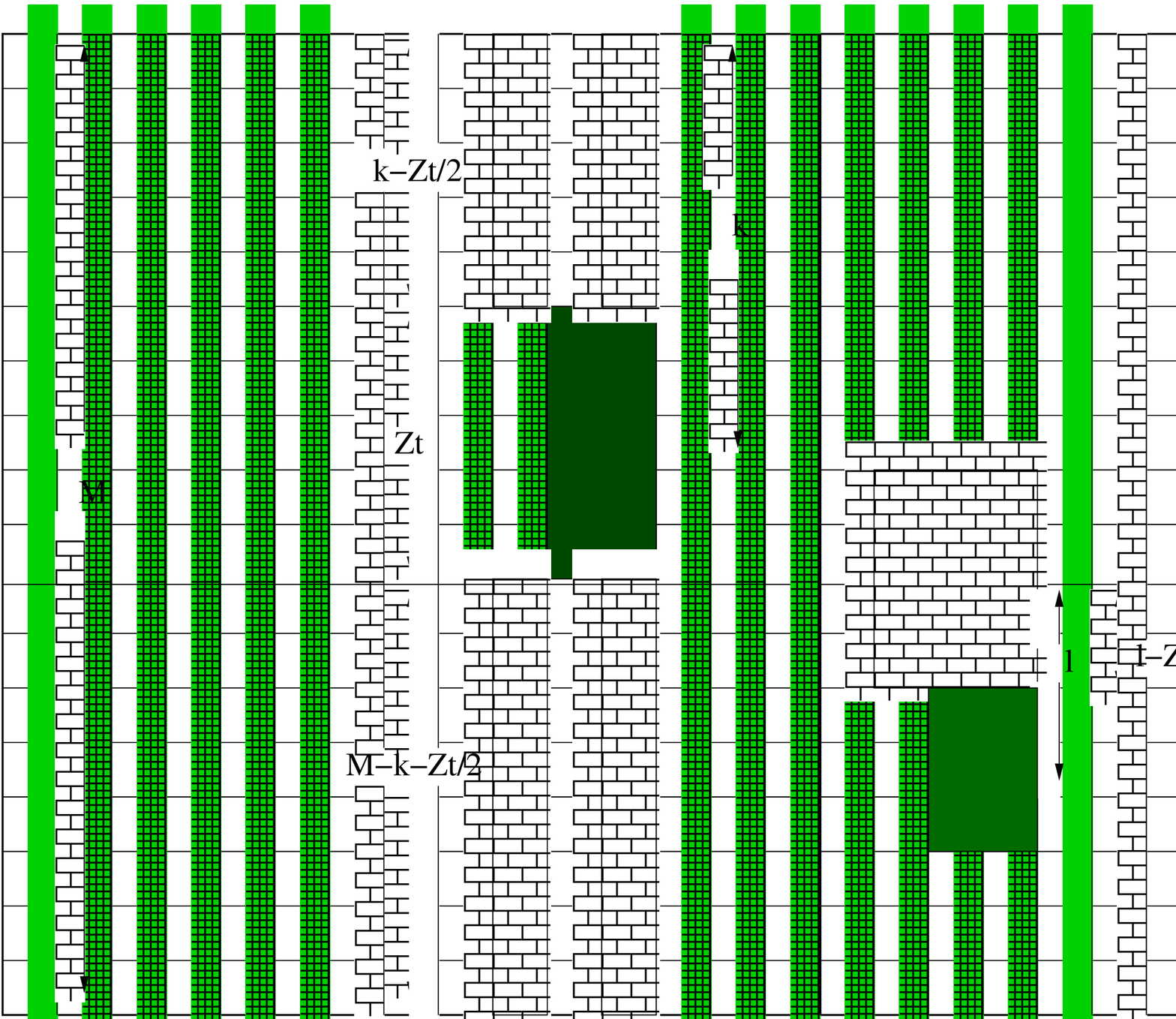}
  \vspace{1ex}
  \caption{Left : Scenario after a time $t_i = t (< t_a)$ after the first nucleation on an empty lattice. The nucleation occured $k$ lattice sites from the nearest horizontal edge. The doubled step (denoted by the shaded region) has grown to a length $Zt$. The sites where a nucleation would lead to frustration ( denoted by the ``walled'' sites ) number $2(M-Zt)$.  Right : A situation after a time $t_i = t$ after a successful nucleation. ``Walled'' sites denote the places where a nucleation could lead to the growing doubled step being ``cut-off'' before it reaches a length $l_0 = l$ on one side. }
  \label{fig:illust}
\end{figure}

We first define $q$ to be the probability that a 
nucleation event leads to a perfectly doubled step on an {\it empty} lattice.
If $I$ is the nucleation rate over the $M \times M$ sample then the probability 
of any one element initiating a nucleation in time $\Delta t$ is
 $I/M^{2} \Delta t$.  Now 
consider  a nucleation event $k$  lattice constants away from the nearest horizontal edge 
of the $M \times M$ lattice (see fig.~\ref{fig:illust}).
 
 Upto a time $t_a = 2k/Z$, both ends of the 
doubled step grow by zipping. For $t > t_a$ only one end grows since the other 
end has already reached the edge of the lattice. This part of the process takes 
a time $t_b= 2(M-2k)/Z$. For  $t < t_a$ the number of 
sites where a nucleation can lead to frustration is $2(M-Zt)$.   For  $ t_a+t_b 
> t > t_a$ the number of 
sites where a nucleation can lead to frustration is $2(M-2k-Z(t-t_a)/2)$. Now 
the probability  at a time $t_i < t_a$ past a nucleation that the doubling steps 
do not  get frustrated in the next $\Delta t$ of time is given by
\begin{align}
  P_a (i) & = ( 1 - \frac{I}{M^2} \Delta t )^{2(M-Zt_i)} \\
     & \approx  \exp \left[- \frac{I}{M^2} 2\left( M-Zt_i \right) \Delta t \right]
\label{eq:qka}
\end{align}
Similarly for $ t_a+t_b > t_j > t_a$, we get
\begin{align}
  P_b (j) & = ( 1 - \frac{I}{M^2} \Delta t )^{2(M-2k-Z(t_j-t_a)/2)} \\
     & \approx  \exp \left[ - \frac{I}{M^2} 2\left( M-2k-Z\left( t_j-t_a \right)/2 \right) \Delta t \right]
\label{eq:qkb}
\end{align}
  Thus the total probability that the nucleation event, $k$ lattice constants 
from the edge,  leads to a perfectly 
doubled step is
\begin{align}
  q_k & \approx \prod_{i} P_a (i) \prod_{j} P_b (j) \\
     & \approx \prod_{i} \exp \left[- \frac{I}{M^2} 2\left( M-Zt_{i} \right) \Delta t_{i} \right] 
\prod_{j} \exp \left[ - \frac{I}{M^2} 2\left( M-2k-Z\left( t_j-t_a \right)/2\right) \Delta t_j \right] \\
     & =  \exp \left[- \int_{t=0}^{t=t_a} \frac{I}{M^2} 2\left(M-Zt\right) dt - \int_{t=0}^{t=t_b} \frac{I}{M^2} 2\left(M-2k-Zt/2\right) dt \right] \\
   & = \exp \left[-\frac{2I}{Z}\right] \exp \left[-\frac{4I}{Z} \left(\left(\frac{k}{M}\right)^2- \left(\frac{k}{M}\right)\right)\right]
\label{eq:qk}
\end{align}
 It is equally probable for the first nucleation to occur at any site along a 
column. We therefore have to average over all possible values of $k$ which are 
integers from zero
to $M/2$. This gives
\begin{align}
   q & = \langle q_k \rangle_k \\
       & = \exp \left[-\frac{2I}{Z}\right] \frac{2}{M} \sum_{k=0}^{k=M/2} \exp \left[-
\frac{4I}{Z} \left(\left(\frac{k}{M}\right)^2-\left(\frac{k}{M}\right)\right)\right] \\
         &  \approx \exp \left[-\frac{2I}{Z}\right] \left( 2 \int_{0}^{1/2} 
\exp \left[-\frac{4I}{Z}\left(x^2-x\right)\right] dx\right) \\
       & = \exp \left[-\frac{I}{Z}\right]\left(\frac{Z}{I}\right)^{\frac{1}{2}}\frac{ 
\pi^{\frac{1}{2}}}{2} \mathrm{erf} \left[\left(\frac{I}{Z}\right)^{\frac{1}{2}}\right]
\label{eq:full-latt}
\end{align}

 Thus we have a relation between the probability of a nucleation leading to a 
perfectly doubled step and the ratio of the nucleation and zippering rates. It 
is to be noted that to preserve invariance under rescaling time, $q$ can only 
depend on the ratio of the rates and not on their absolute magnitude. Happily, equation \ref{eq:full-latt} respects this invariance. \par
We now return to the question of when the approximation of considering only order-one defects is valid. To do this, we compute the value of the mean free path of a successful nucleation, defined earlier, explicitly. We first compute 
the probability that the doubled step length on one side of the point of 
nucleation exceeds $l_0$. At any 
time $t_i$ after the nucleation event the probability that the doubling step 
does not get stopped by doubling events in its path in the next time step ($\Delta t$) is given by
\begin{align}
  P (i) & = ( 1 - \frac{I}{M^2} \Delta t )^{6(l_0-Zt_i/2)} \\
     & \approx  \exp \left[- \frac{I}{M^2} 6\left(l_0-Zt_i/2\right) \Delta t \right]
\end{align}
Here we take the probability of not stopping the zippering end of the 
doubling step in the next $\Delta t$ of time to be the probability of having no 
nucleation in time $\Delta t$ at any site raised to the power of the number of sites 
where such a nucleation would lead to stoppage (see fig.~\ref{fig:illust}). Now the total probability is 
simply the product of the $P_{i}$'s which can be written as the exponential of 
an integral yielding
\begin{equation}
   P(l>l_0) = \exp \left[ -\frac{6I}{Z} \left(\frac{l_0}{M}\right)^2 \right]
\end{equation}
Knowing this probability distribution we can calculate the mean length $\langle 
l \rangle$, which is what we defined as the mean free path. The mean free path is 
thus
\begin{equation}
  \langle l \rangle =  2 \int_{0}^{\infty} \frac{dP}{dl} l dl = M  (\frac{4 \pi}{6})^{\frac{1}{2}} (\frac{Z}{I})^{\frac{1}{2}}
\end{equation}
 Using the explicit formula for $ \langle 
l \rangle $ in terms of $I/Z$ above, the condition for neglecting higher order 
defects reduces to $I/Z \ll 4 \pi /6 \sim 2.09$. As noted before for most 
realistic experimental situations this situation is easily satisfied.
Simulation results presented later support our assumption.\par
Now, the value of $q$ derived above neglects the influence of nucleations other than the ones that can frustrate the initial zippering double step.  For example a nucleation directly below and aligned with the original growing double step will block sites where nucleations could have frustrated the original growing double step. This is also true for nucleations on columns on either side of the original doubling step. In general a nucleation occuring many columns away may still influence the probability of the original nucleation forming a perfectly doubled step. Intuitively we would expect that the further such a column is from the original nucleation, the less its influence will be. Indeed for a nucleation occuring $j$ columns away the probablity that its effect will propagate to the column where the original nucleation took place will go as $(I/Z)^j$.  This is because, for this to happen, we need $j$ nucleations (for each intervening column), each of which will occur roughly with a probability proportional to $I/Z$. Thus for small enough values of $I/Z$ we may neglect the effect of columns that are further away. However to be sure that this is not a big effect we need to ascertain the effect of nucleations on columns that are closest to the original doubling step. To do this, we need to take into account the probability that a site at which a nucleation could lead to frustration may not be available for occupation. Thus equation \ref{eq:qka} will read
 \begin{equation}
  P_a (i)  = \left( 1 - \frac{I}{M^2} u \Delta t \right)^{2(M-Zt_i)}
\end{equation}
The extra factor $u$ reflects the probability that the site in question is available for occupation. $u$ may be roughly related in a ``mean field'' sense to the average fraction of unoccupied sites at that time. As an approximation we take this fraction to be $P_1$ for the case with an infinite zippering rate. Then going through the above analysis will yield a ``corrected'' value of $q$. We find that $q$ goes up by a factor which is about $6.2\%$ when $q\sim0.5$ and less than $0.1\%$ when $q\sim0.9$. Thus we find that neglecting these effects does not alter our results by much.\par
It is to be noted that when the only structures 
present were perfectly doubled steps ($Z = \infty$), the mapping to the RSA problem was exact. Now we have two kinds of entities. The doubled steps still cannot be overlapped
and are put down randomly. They are still amenable to RSA analysis. However the frustrated dead ends can become defects of order two (fig.\ref{fig:defect}) when the sections of them that are still single steps merge with a neighboring step if available. If this occurs during the process when
there are still pairs of single steps that can form perfectly doubled steps, it will 
affect the dynamics and hence the asymptotic fraction of different species. It is this process that we neglect as a first approximation. Later simulation results show that
this is a reasonable assumption.\par
 We now consider how the competition parameter $q$ is altered when the surrounding lattice is not empty as discussed above. The value of $q$ does depend on the environment in which the nucleation occurs. We can imagine three possible scenarios. (1) Columns on either side of the freshly nucleated double step are already occupied (doubled). In this case $q=1$, since this nucleation cannot be frustrated. (2) Columns on both sides are empty (single). In this case we may use the result for the empty lattice, since for low enough $I/Z$, as we argued before, the state of columns further away will not matter much. (3) One side of the freshly nucleated  double step is occupied. The correct value to be used here will be $q^{1/2}$. The relative probabilities of these scenarios occuring will evolve as a function of time. In order to account for this we ought to use the more general treatment, where $q$ can be an arbitrary function of time, presented in the previous section. However since most of the defects form in the early stages, we anticipate that the asymptotic fraction of defects will not be altered significantly if we use a constant value of $q$, evaluated for the empty lattice. Our simulation results confirm this view. The dynamics will however will be sensitive to changes in $q$ and this will be addressed in the section on time dependence. For now we take $q$ to be a fixed constant throughout the process.
\par
\begin{figure}
  \centering
  \includegraphics[width=4.0in]{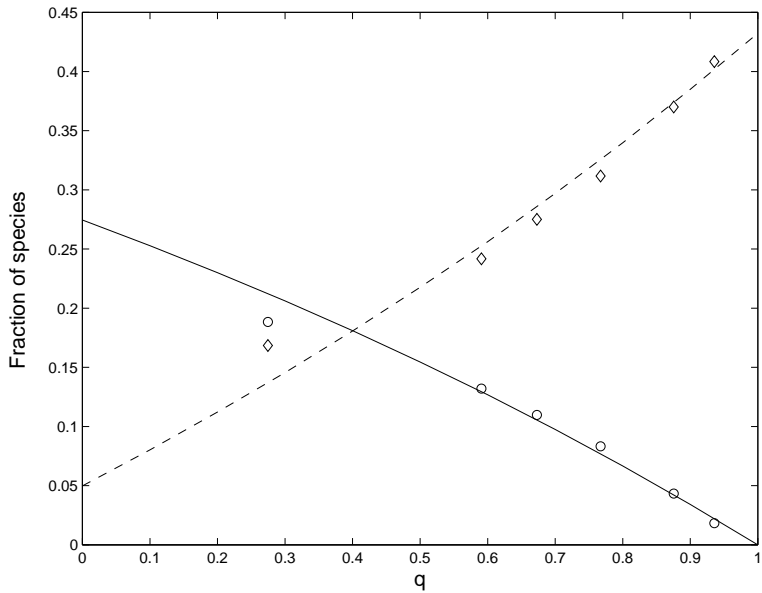}
  \vspace{1ex}
  \caption{Plots of fraction of doublets and triplets as a function of $q$. 
The fraction of a species is the number of that species divided by the 
size of the lattice (60 here). Solid and dashed curves are obtained by numerical 
integration of \ref{eq:asymfrac} and refer to the fraction of frustrated dead 
ends and perfectly doubled steps respectively. The simulation results for the 
fraction of frustrated dead ends (open circles) and perfectly doubled steps 
(diamonds) are also shown and are plotted with $q$ values computed using the 
equation (\ref{eq:full-latt})}
  \label{fig:fig2}
\end{figure}

 
We can now view the process as a RSA of a binary mixture of dimers and trimers 
on a 1-D lattice. A dimer adsorbs onto the lattice with a probability $q$ and a 
trimer with a probability $1-q$. If $k_{0}$ is the overall attempt rate at which 
nucleations are tried on the 1-D lattice then the dimers have an attempt rate  
$k=k_{0} q$ and the trimers have an attempt rate $k' = k_{0} (1-q)$. 
We now use the results derived for the process on an infinite 1-D lattice with dimers attempting 
to adsorb at a rate $k$ and trimers at a rate $k'$ (as in previous section). Using the fact that here $k$ and $k'$ are constants eq.\ref {eq:P2t} yields
 \begin{equation}
   P_{2} =  \exp \left[-kt + 2 e^{-(k+k')t} +\frac{k'}{k+k'} e^{-2(k+k')t}  -
(2+\frac{k'}{k+k'})\right]
\end{equation}
and hence from (\ref{eq:ansatzt}) an explicit solution for all $P_{n}$ for $n \ge 
2$. For $n=1$ we have
 \begin{equation}
  \frac{dP_{1}}{dt} = -2kP_{2} - 3k'P_{3}
  \end{equation}
which upon integration gives
 \begin{equation}
  1 - P_{1}(t=\infty) = \int_{0}^{\infty}( 2kP_{2} +3k'P_{3} ) dt
 \label{eq:asymfrac}
 \end{equation}
$P_{1}(t=\infty)$ is simply the fraction of space occupied by isolated dead ends 
at asymptotic coverage. The two terms on the right are simply the fraction of 
space occupied by the dimers and trimers or double steps and frustrated dead 
ends respectively. Figure~\ref{fig:fig2} shows a plot of the fraction of various 
species as a 
function of $q$. This is the only parameter in the problem as the 
asymptotic coverages are clearly independent of the overall rate $k_{0}$. Thus 
we have demonstrated that both the nature and number of defects in the final 
state
depend only on the dimensionless parameter $q$. It is 
to be noted that the isolated single step population (not shown) does not change 
much
 ($ 8-13.5 \%$) with 
$q$ in contrast to double steps and frustrated dead ends. This is in qualitative 
agreement with experimental results \cite{yi}. It is also to be noted that when 
$q \lesssim 0.58$
, the fraction of width occupied by the defects exceeds that 
occupied by
 perfectly doubled steps. This characterization can be used 
to infer 
the ratio $I/Z$ by simply looking at the number of defects in the final 
structure. One thus gains information about the dynamics of the process from the 
final state. \par

\subsection{ Scaling of the fraction of defects}
A point of interest is the case where $Z \gg I$. Here the nucleations are 
the rate limiting step and the zippering occurs almost instantaneously. In this 
case 
$k' \rightarrow 0$ and we retrieve the $Z = \infty$ case where we can explicitly
 solve for $P_{1}$ yielding 
$P_{1}(t=\infty)= e^{-2}$. Thus even in this situation $13.5 \%$ of the steps 
remain 
as isolated step defects.\par

\begin{figure}
  \centering
  \includegraphics[width=4.0in]{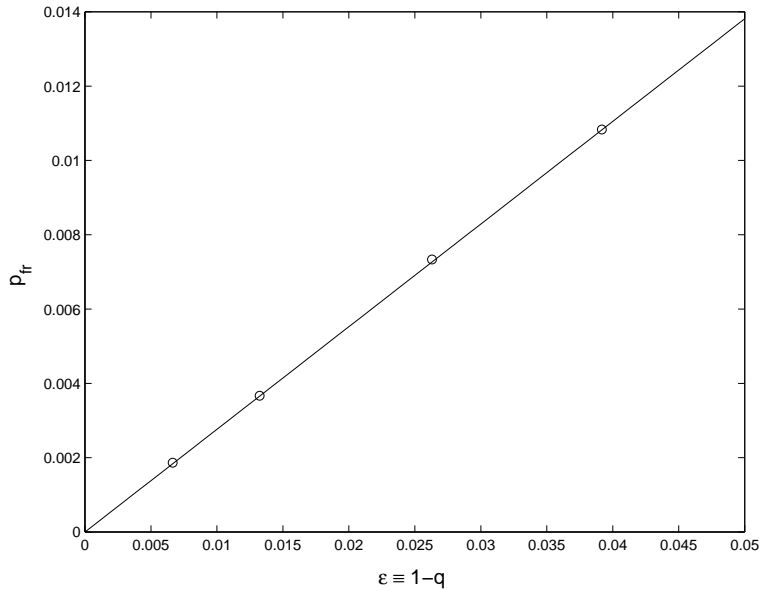}
  \vspace{1ex}
  \caption{Plot of fraction of frustrated dead ends versus $\epsilon \equiv 1-q$ 
showing the linear relation.  The open circles are simulation results and the 
straight line is a best fit to these points. The best fit line has a slope of 
$0.28 \pm 0.01$ and an intercept of $0.0000 \pm 0.0002$.}
  \label{fig:q1fig}
\end{figure}

In the limit of $q \rightarrow 1$ the mapping to the RSA problem becomes exact. 
We now 
consider the scaling of the number of frustrated dead ends with $\epsilon \equiv 
1-q$ in this limit. The fraction of frustrated dead ends (defects of order one) 
is given by
\begin{align}
  p_{fr} & = \int_{0}^{\infty} k' P_3 dt \\
          & = \int_{0}^{\infty} k_0 (1-q) P_3 dt \\
          & =  k_0 \epsilon \int_{0}^{\infty}  \exp [-k_0 t] P_2 dt \\
          & =  k_0 \epsilon \int_{0}^{\infty}  \exp [-k_0 t] \exp \left[ -kt + 
2 e^{-(k+k')t} +\frac{k'}{k+k'} e^{-2(k+k')t}  - (2+\frac{k'}{k+k'}) \right] dt 
\end{align}
Here $k=k_0 (1-\epsilon)$ and $k'=k_0 \epsilon$. Using these in the above 
expression and retaining terms to lowest order in $\epsilon$ we obtain

\begin{align}
   p_{fr}& = k_0 \epsilon e^{-2} \int_{0}^{\infty} \exp [- 2 k_0 t+ 2 
e^{- k_0 t}] dt \\
      & =  \epsilon e^{-2} \int_{0}^{\infty} \exp [- 2 x+ 2 e^{- x}] dx 
\\
      & = 0.284... \epsilon
\end{align}
Thus for small $\epsilon$ the fraction of frustrated dead ends  rises linearly 
with $\epsilon$ with a slope of $0.284$. Figure~\ref{fig:q1fig} shows a plot of 
values of $p_{fr}$ obtained from the simulations described in the section.III  
versus $\epsilon$. The error bars on the individual data points are about $5\%$. 
The line plotted is a best fit line to the points and yields a slope of $0.28 
\pm 0.01$ and an intercept of $0.0000 \pm 0.0002$. This agrees very well with 
our prediction and shows that $p_{fr}$ indeed rises linearly with $\epsilon$ 
with a slope of $0.28$. The observed number of frustrated dead ends thus gives us explicit information about the competition parameter $q$  and hence the experimental rates.\par
\par
\subsection{Accounting for nucleations outside the field of view}

In the above analysis we assumed that {\it all} doubling occurred as a result of 
nucleations that occurred within the region of observation. Clearly this need not 
be the case. Nucleations that occur outside the region of observation can lead to 
doubled steps within this region. In a typical experimental situation for 
example one would be most likely to examine some subsection of the sample and 
take that subsection as being representative. We thus need to account for unobserved nucleations in 
order to make contact with experiment.\par
\begin{figure}
  \centering
  \includegraphics[width=4.0in]{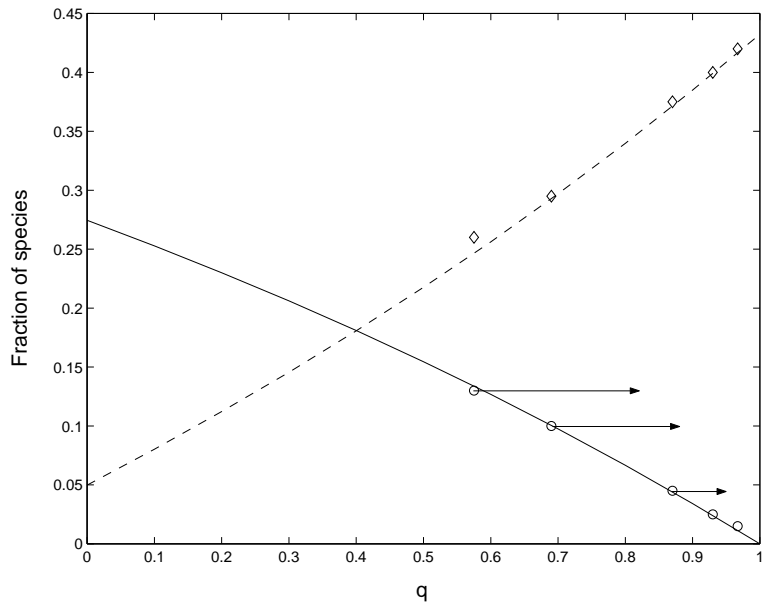}
  \vspace{1ex}
  \caption{Plots of fraction of doublets and triplets as a function of $q$. 
Solid and dashed
 curves are theoretical curves and refer 
to the fraction of frustrated dead ends and perfectly doubled steps 
respectively. These are identical to the curves in fig.~\ref{fig:fig2}  The
simulation
 results for the fraction of frustrated dead ends
(open circles) and perfectly doubled steps (diamonds) are also shown but with
$q$ being
 computed using the relation for the case of an open subsystem. The arrows on selected data points indicate how much they would shift if $q$ were calculated using the relation for the case of a closed system. The
fraction of a species is the number of that species divided by the
size of the lattice (20 here).}
  \label{fig:fig3}
\end{figure}
\begin{figure}
  \centering
  \includegraphics[width=4.0in]{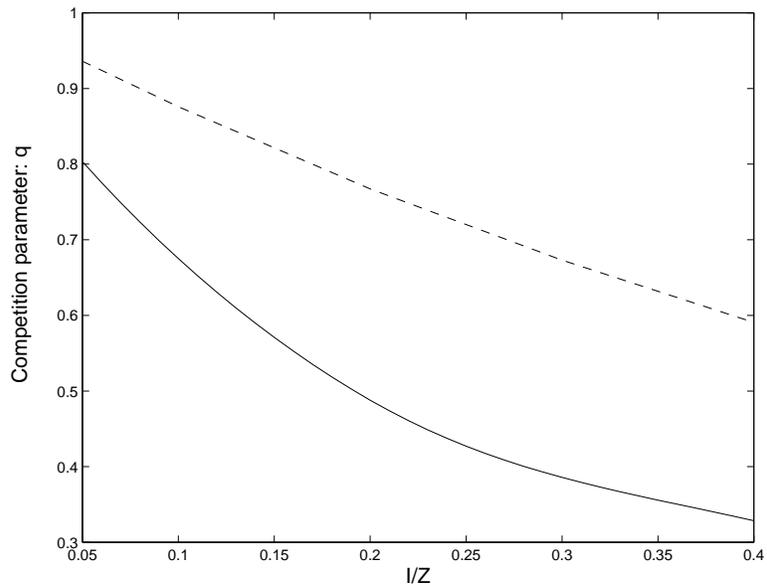}
  \vspace{1ex}
  \caption{Competition parameter versus $I/Z$ for the closed system (dashed line) and the open subsystem (solid line). Values of $q$ are lower for the open subsystem case for the same values of $I/Z$.}
  \label{fig:submat}
\end{figure}

Most of the analysis done before still holds for this case. Nucleations occurring outside the field of view will increase the overall attempt rate (since there will be doubled steps formed by nucleations occurring outside and zippering in). They will also make it more probable for a given nucleation to be frustrated (since nucleations occurring outside can also frustrate doubling steps within the field of view). If we are interested only in the asymptotic fraction of defects, we do not need to consider changes in the overall attempt rate. The only difference 
will arise in computing the probability that a given attempt at a doubled step 
succeeds without being frustrated i.e. our competition parameter $q$. We take the region of observation to be an $M 
\times 
M$ lattice embedded in a larger lattice which is $ 3M \times 3M$. The 
 choice is simply made based on convenience and for ease of comparison with 
simulations. First we recall the concept of a mean free path for a zippering 
double step. We defined this to be the average length to which a zippering double 
step can grow from the nucleation point without being hindered by a nucleation 
in its path. Taking into account the probability per unit time of a nucleation 
occuring in the path and the zippering speed it was shown that the 
probability of a zippering double step growing to a length {\it greater} than 
$l_0$ is
\begin{equation}
   p(l>l_0) = \exp \left[ -\frac{6I}{Z} (\frac{l_0}{M})^2 \right]
\label{eq:pl}
\end{equation}
where $I$ is the nucleation rate over an $M \times M$ area. Knowing this 
probability distribution we can calculate the mean length $\langle l \rangle$ 
which is what we defined as the mean free path. The mean free path is thus
\begin{equation}
  \langle l \rangle =  M \left(\frac{4 \pi}{6}\right)^{\frac{1}{2}} \left(\frac{Z}{I}\right)^{\frac{1}{2}}
\end{equation}
Now we assume that the {\it only} nucleations that will have an effect on the 
region of observation  occur either in this region or $\langle l \rangle$ above or 
below this region. This is because a zipper originating from a nucleation farther 
away is likely to be cut off by nucleations in its path. 
 As before we consider the scenario at a 
time 
$t$ after a nucleation event  inside the region of interest, $k$  lattice 
constants away from the nearest edge of the $M \times M$ lattice on an otherwise 
empty lattice. The probability that this 
nucleation leads to a perfectly doubled step may be derived in a manner analogous to the way we derived equation (\ref{eq:qk}). There are two factors contributing to this probability. First there is the probability of being frustrated by nucleations inside the field of view. This is exactly the same as before. We denote this as $P_{in}$, which is given by
\begin{equation}
  P_{in}  = \exp \left[-\frac{2I}{Z}\right] \exp \left[-\frac{4I}{Z} 
\left((\frac{k}{M})^2-(\frac{k}{M})\right)\right]
\end{equation}
This follows from eq.\ref{eq:qk}. The second factor  which we now consider, is to account for nucleations occurring outside the field of view. Now 
the probability  at a time $t_i < t_a$ past a nucleation that the doubling steps 
do not  get frustrated in the next $\Delta t$ of time is given by
\begin{equation}
  P^1_{out} (i)  = \left( 1 - \frac{I}{M^2}\langle l \rangle P_2 \langle p \rangle \Delta t \right)^{4} 
\end{equation}
Here $t_a=2k/Z$ as before. $\langle p \rangle$ refers to the average probability that a nucleation within a mean free path distance of the horizontal edge of the field of view, zippers to the edge without being cut off. This is simply the average of the probability given by eq.\ref{eq:pl} over this region.
\begin{equation}
 \langle p \rangle = \left(\frac{Z}{6I}\right)^{1/2} \frac{M}{\langle l \rangle} \frac{\pi^{1/2}}{2} {\mathrm{erf}} \left[  \left(\frac{Z}{6I}\right)^{-1/2} \frac{\langle l \rangle}{M} \right]
\end{equation}
$P_2$ accounts for the probability that a nucleation is indeed possible at the site in question. We simply choose for $P_2$ the solution for the infinite zippering rate case (eqs.\ref{eq:infsol} and \ref{eq:ansatz0}). The factor of $4$ comes from the fact that there are four columns along which a nucleation from outside can zipper in to frustrate the original nucleation (see fig.\ref{fig:illust}).
 We can write down a similar expression for times $t_a<t_i<t_a+t_b$ where $t_b = 2(M-2k)/Z$.
\begin{equation}
  P^2_{out} (i)  = \left( 1 - \frac{I}{M^2}\langle l \rangle P_2 \langle p \rangle \Delta t \right)^{2} 
\end{equation}
The change here is in the exponent which changes from $4$ to $2$, since once one end has reached the edge, there are only two columns left along which a nucleation from outside can zipper in to cause frustration. We now take the product $P^1_{out}P^2_{out}$ and integrate over time to get
\begin{equation}
  P_{out}  = \exp \left[ \frac{-4I}{M^2}\langle l \rangle \langle p \rangle \int_0^{t_a} P_2 dt \right] \exp \left[ \frac{-2I}{M^2}\langle l \rangle \langle p \rangle \int_0^{t_b} P_2 dt \right]
 \label{eq:pout}
\end{equation}
 We similarly consider  the probability that a
nucleation event that occurs outside the region (within $\langle l \rangle$ of 
the edge of the region of interest) results in a perfectly doubled step.
\begin{equation}
  P^{0}  = \exp \left[ -\frac{2I}{Z} \right] \exp \left[ \frac{-2I}{M^2}\langle l \rangle \langle p \rangle \int_0^{2M/Z} P_2 dt \right] 
\end{equation}
This comes from putting $k=0$ in the product $P_{in}P_{out}$.
 This has to be weighted by the probability that the particular 
nucleation results in a zipper that makes it to the edge. This probability as 
calculated before (eq.\ref{eq:pl}) is
\begin{equation}
   p(j) = \exp [ -\frac{6I}{Z} (\frac{j}{M})^2 ]
\end{equation}
Now we simply average  the probability of getting a perfectly doubled step over 
all positions (both inside and outside) with the appropriate weights following the steps to get eq.\ref{eq:full-latt}. It is to be noted that if the distance from the edge of the field of view to the edge of the outer boundary is less than the mean free path, then $\langle l \rangle$ in the above expressions is to be replaced by this distance. The integrals cannot be performed to yield a closed form answer. These however can be numerically evaluated for specified values of $I/Z$.
 The 
dependence
of the fraction of species on $q$ remains the same as before. Only the relation 
between $q$
and $I/Z$ has changed. Figure~\ref{fig:fig3} shows a plot of the fraction of 
various species
 as a 
function of $q$. It is to be noted that, for the same value of $I/Z$, an open
 subsystem 
will have a smaller value of $q$ and hence a larger number of frustrated dead 
ends than
a closed system of the same size (see fig.\ref{fig:submat}). This is what one would intuitively expect.
 \par
One can use this analysis to infer the number and nature of defects in a larger 
sample
simply by looking at a small patch and inferring the value of $I/Z$. However 
care must
be taken if the calculated value of $I/Z$ for the larger sample  exceeds about 
0.5. 
One must then do an extended analysis incorporating a finite number of higher 
order 
defects dictated 
by the value of $I/Z$. It is to be noted that this treatment contains several ad hoc approximations and should only be taken as a rough estimate and a proposed methodology.

\subsection{Time Dependence}

Until now we have been looking at the asymptotic ($t \rightarrow \infty$) 
limit of the surface morphology. In this non-equilibrium problem however the 
final morphology is dictated by the dynamic evolution and hence it is 
interesting to look at this evolution. Experimentally too there has been recently 
much progress in studying the time evolution of the step doubling process 
\cite{pearl1,pearl2,niu,hk}. \par
We first introduce an order parameter $\psi$ which  we define to be the fraction 
of sites that have undergone doubling.  A completely empty matrix in which 
all steps are 
single corresponds to $\psi=0$ and a completely full state in which all sites 
are doubled corresponds to $\psi=1$. At the beginning ($t=0$) we start with a 
situation where all the  steps are single ($\psi=0$). Now consider the situation 
after a time $t$. We define the average time it takes for a perfectly doubled 
step to 
form to be $t_{2}^{*} = 3M/2Z$ . We assume it takes roughly half the 
time for a frustrated dead end to form and define $t_{3}^{*} = 3M/4Z$.  We make these approximate estimates since we would just like to give a qualitative sense of the time dependence. Now all 
double steps initiated before time $t-t_{2}^{*}$ and all frustrated dead ends 
initiated before time $t-t_{3}^{*}$ have  formed completely. Entities initiated 
within this interval have formed partly. So we have

 \begin{equation}
  \psi (t)  = \int_{0}^{t-t_{2}^{*}} 2kP_{2} dt' + \int_{0}^{t-t_{3}^{*}} 
2k'P_{3} dt'
            + \int_{t-t_{2}^{*}}^{t} 2k'P_{2} \frac{t-t'}{t_{2}^{*}} dt'
             + \int_{t-t_{3}^{*}}^{t} 2k'P_{3} \frac{t-t'}{t_{3}^{*}} dt'
 \label{eq:dynamics}
 \end{equation} 

\begin{figure}
  \centering
  \includegraphics[width=4.0in]{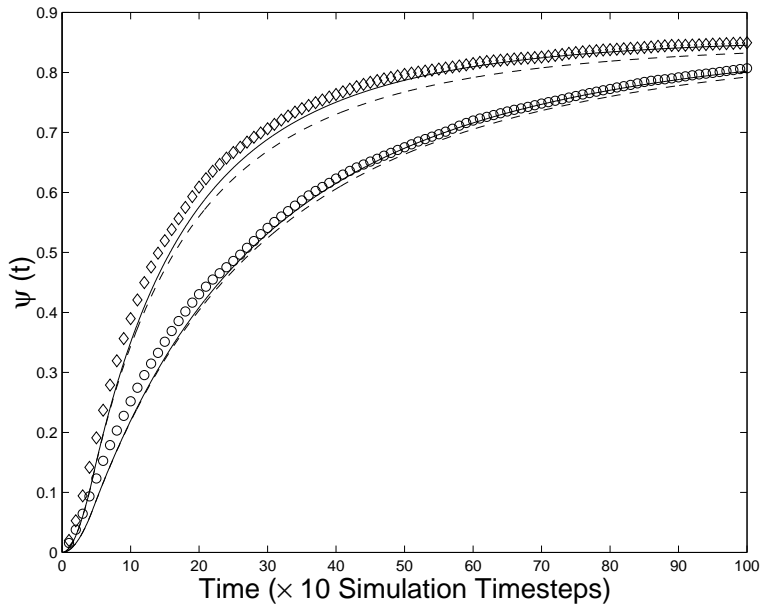}
  \vspace{1ex}
  \caption{Plots of $\psi (t)$ versus time for two different values of $I/Z$ 
(0.05 (lower curve) and 0.1 (upper curve)). The solid curves are obtained by 
numerical evaluation of \ref{eq:dynamics} for  the case with a time dependent 
competition parameter $q(t)$. Dashed curves are obtained by 
numerical evaluation of \ref{eq:dynamics} for  the case with a time independent 
competition parameter $q_0$. Simulation results obtained for 
$I/Z=0.05$ (open circles) and $I/Z=0.1$ (diamonds) are also shown.}
  \label{fig:fig4}
\end{figure}

The first two terms account for the completely formed (completed double steps 
and frustrated dead ends) structures. The third and fourth terms account for the 
structures that are partly formed. The factor of two in all the terms comes from 
the fact that each doubled step occupies two columns. The frustrated dead ends 
also effectively occupy two columns (in terms of area) though they span three columns. Since we 
have explicit solutions for $P_{2}$ and $P_{3}$ as functions of time we can 
compute $\psi (t)$. The only parameter we have not yet evaluated is the overall 
attempt rate $k_0$. We may think of each doubled step as being caused by 
one nucleation and each frustrated dead end being caused by two. Then the 
attempt rate of dimers plus twice the attempt rate of triplets ought to be equal 
to the nucleation rate per column.
  \begin{align}
   k + 2k'  & = k_0 q + 2 k_0 ( 1-q ) \\
     & = k_0 ( 2-q ) = \frac{I}{M}
\end{align}
 which yields
  \begin{equation}
      k_0 = \frac{I}{M ( 2-q )}
\label{eq:koft}
  \end{equation}
Having evaluated $k_0$ we can now compute $\psi (t)$ explicitly in terms of $I$,
$Z$ and $M$. Figure~\ref{fig:fig4} shows a plot of the computed $\psi (t)$ (dashed line) as a 
function of time 
for two different values of $I/Z$ (0.05 and 0.1). $Z=2$ in both cases. The 
difference between the two curves shows that the dynamics is quite sensitive to 
the parameters. Thus combining the information about the defects in the 
asymptotic structure and a measurement of the time dependence of $\psi (t)$
 will allow us to uniquely determine both the nucleation and zippering 
rates. The mismatch between the analytical and simulation curves at late times
 was anticipated
before and comes
from the assumption of a constant $q$.
To fix this, we use the solution for the case with the time dependent competition parameter, $q(t)$. To do this, we need to compute $q(t)$. Now, the value of $q$ depends on where a particular nucleation event takes place. As mentioned before, if the nucleation takes place at a site which belongs to a set of two empty sites with doubled steps on either side, the value of $q$ is unity since this doubling step cannot be frustrated. This situation would occur with a probability $1-P_3$. This includes the probability of picking isolated single steps where nucleations are not possible, resulting in the attempt being rejected. As a first approximation we assume that nucleation attempts at all other places (with probability $P_3$) have the value of $q$ computed for an empty lattice, $q_0$. In principle we can systematically refine this approximation by considering sites adjacent to a double step but belonging to a large set of consecutive empty sites having a different value of $q$ and so on. However we find that the first level of approximation is sufficient for our purposes. Thus we take the competition parameter to have a time dependent value
\begin{equation}
  q(t) = q_0 \times ( P_3 ) + 1 \times (1-P_3)
\label{eq:qoft}
\end{equation}
We now solve the set of equations (\ref{eq:P2t}), (\ref{eq:qoft}) and (\ref{eq:koft})
self-consistently by an iteration procedure starting with the solution for the case 
with a constant competition parameter, $q_0$. The results after a couple of iterations are plotted in ~\ref{fig:fig4} as solid lines. One immediately sees that these agree with the simulation points much better at later times. Thus our treatment allows us to capture the time evolution of the surface morphology fairly accurately.

\section{Simulation}
 
 We now describe the simulation that produced the data in figures 2--5.  This simulation uses the coarse-grained representation of figure 1 but makes none of the assumptions leading to the curves in figures 2--5.  It thus serves as a test of our RSA approximations.   We use an $M \times M$ 
matrix to mimic a square section of the sample. Every column of the matrix is 
regarded as a step. We have two experimental parameters that we may use: the 
nucleation rate $I$ and the zippering rate $Z$.
 A typical simulation cycle is as 
follows:\\
(1)A randomly chosen site of the matrix is addressed.\\
(2) If a chosen site is unoccupied and its neighbor to the right is also unoccupied 
these two sites are allowed to combine to form a nucleus with a probability 
$p=I/M^{2}Z$. This takes care of the relative rates of nucleation and zippering. 
Now  these two sites are considered occupied.\\
(3) Steps (1) and (2) are repeated making sure sites are not addressed twice until 
all sites of the matrix have been addressed.\\
(4) All  steps that have previously initated doubling are allowed to grow by one 
lattice unit at each free end, if possible.\\
(5) Steps (1)-(4) are repeated until no new nucleations are possible and the surface 
is comprised only of perfectly doubled steps, frustrated dead ends and isolated 
single steps.\\
(5) The results are averaged over hundreds of runs for different values of the ratio 
$I/Z$.\\
\par
\subsection{Final State}
The size of the matrix used was $M=60$ since the STM experiment described in  
\cite{yi}
has a field of view roughly 60 terraces wide and 60 terraces tall. 
In the range of parameter values we looked at the number of defects 
that were not simple frustrated dead ends were few ($\sim 15\%$ of the defects 
 for $q=0.5$ and decreasing with increasing $q$), thus 
justifying the rationale for our theoretical assumption. When these were 
encountered they were decomposed into constituent dead ends and counted as that 
many effective frustrated dead ends. For example an order two defect was
counted as one frustrated dead end and one isolated step. 
Figure~\ref{fig:fig2} plots the results of 
the simulation 
for  various values of $I/Z$. We see very good agreement between the simulation 
and values predicted by theory for small $I/Z$. We also see that the theory 
begins to break down for large values of $I/Z$ ($>0.46$). It is to be noted that 
the largest number of perfectly doubled steps 
occurs in the limit of $Z \gg I$ and is about $43.75 \%$. This is consistent with
our prediction for the $Z = \infty$ case. \par
We also consider the case where the region of interest is part of a larger 
region. We utilize the same $M=60$  matrix and we now count the defects and 
doubled steps in the central $20 \times 20$ submatrix. Figure~\ref{fig:fig3} 
plots the results 
of the simulation for  various values of $I/Z$. The agreement between the 
theoretical curve and the simulation values appear to be quite good in this case 
too. 
\subsection{Time dependence}
We also look at the evolution of the order parameter $\psi$ during a simulation 
run. The simulation is done on a $60 \times 60$ matrix as before. The fraction 
of sites that are occupied ( doubled ) is registered after 
every 10 simulation timesteps. This data is recorded for hundreds of whole 
simulation runs. For each block of 10 timesteps we then record the average 
$\psi$ over all the runs. Figure~\ref{fig:fig4} shows a plot of the averaged 
evolution of the 
order parameter as a function of simulation time for two different values of 
$I/Z$. We see that the simulation data and the theoretical curves agree quite 
well. It is to be noted that there are no adjustable parameters. \par
Thus overall the simulation results are in agreement with the theory and the 
theory effectively captures both the dynamics and the details of the asymptotic 
surface morphology.

\section{Discussion of Results}
  We now compare our results to experimental data so as to be able  draw some 
physical conclusions. In the experiments by Wang {\it et al}\cite{yi} it was 
noted
 that under certain optimal 
conditions, a 100 nm by 100 nm section of the sample exhibited 5-6 frustrated 
dead ends after structural evolution had reached an asymptotic stage. The 
step zippering rate was measured to be 3.7 \AA $s^{-1}$. The case to which this 
data ought to be compared is the one where the region of interest is embedded in 
a larger region. From figure~\ref{fig:fig3} we immediately see that to 
get 5-6 frustrated dead ends in a $60 \times 60$ matrix we require $q \approx 
0.7$. 
This gives us a value for the ratio $I/Z \approx 0.12$.  
 Knowing the experimentally measured zippering rate we can also deduce the true 
nucleation rate. In our model we have $Z=2$ measured in units of step width per 
simulation timestep. The step width is about 1.65 nm  which tells us that each 
simulation timestep corresponds to $2 \times 1.65/0.37 \approx 9$ seconds. Thus 
the true nucleation rate in this case would be $0.12/9 = 0.0133$ nucleations per 
second over the 100 nm by 100 nm section of the sample. Our analysis hence helps 
pin down the true experimental parameters simply by looking at the defects in 
the asymptotic stage.\par
Our analysis of the dynamics also gives us more useful insights. In particular
we notice that $P_3 \sim \mathrm{exp} (-kt) P_2$ (from eq.(\ref{eq:ansatzt})).
This tells us that the rate of  defect formation drops exponentially faster than
the  rate at which perfectly doubled steps form. Often one wishes to minimize the
number of defects.
The way to do that would be to have a very low nucleation rate (for a given 
zippering rate
). However then reaching the final state would take a very long time.  Since we 
anticipate
that most of the defects will be formed in the intial stages we could start with 
a low
nucleation rate and after some time jump to a much higher rate. This would mean 
reaching 
the final state much faster with only a small increase in the number of defects.
As an example we ran a simulation with $I/Z=0.05$ on a $M=60$ lattice. The 
number of defects
 was roughly one and it took 3600 timesteps to go to completion. A run where we 
started with
 the same value of $I/Z$ and then switched to a value 20 times higher ($I/Z=1$) 
after
500 timesteps yielded {\it two} defects and took only 700 timesteps to 
complete. In
contrast if we run the simulation for $I/Z=1$ from the beginning we get 
approximately 12 
defects. Thus
we gained a factor of 5 in time for a minor increase in the number of defects. 
One can also imagine trying different time dependent protocols to optimize the 
number
of defects and the time.
\par
Our prediction for the dynamic evolution of the order parameter 
(figure~\ref{fig:fig4}) is qualitatively similar to experimental data by Niu 
{\it et al} \cite{niu}. The authors use a phenomenological approach to fit their 
data using second order rate kinetics. Though the fit is good the analysis 
neglects that the fact that only neighboring steps can double. Another approach 
by Khare  {\it et al} \cite{khare} analyzes the dynamics in terms of first 
passage times of random walkers using a fit with three adjustable parameters. 
However as the authors note, they do not take into account the formation of 
defects like the isolated steps and frustrated dead ends. Our analysis allows us 
to inspect the dynamics with no adjustable parameters if we first extract the 
relevant parameters from an inspection of the asymptotic structure. It also 
takes into account the formation of defects and their effect on the subsequent 
dynamics. This cannot be ignored for a manifestly non-equilibrium problem such 
as this.\par

There are however several details which we have ignored in this analysis. 
Firstly it was noticed \cite{pearl2} that zippering occurs much more 
slowly when the steps are surrounded by already doubled steps. Secondly we have 
not taken into account correlations between neighboring zippers. Another avenue 
of interest would be to integrate the time distribution of nucleation events 
postulated by Khare  {\it et al} \cite{khare} with our RSA analysis. One could
also incorporate higher order defects by a straightforward elaboration of our 
treatment.
Thus it appears that our RSA approach will be applicable to a a range of such propagating surface reconstruction processes. Though we know of only one such process at the moment, many others are sure to emerge as atomic scale knowledge of adsorption on solid surfaces improves.

\section{Conclusion}
We have shown here how emergent features of the novel step-doubling process can be quantitively understood.  Our approximation of the process as  a form of random sequential absorption leads to successful predictions in regions of experimental interest.  The approximation permits simple analysis, yet it shows that naive analysis based on equilibrium statistics is misleading.  The inadequacy of an equilibrium treatment is further apparent in the time-dependence of our  results.  Changing the growth conditions over time can have a striking effect on the final state.  
\par
We have focussed on predicting the incidence of a particular type of defect: the frustrated dead end.  But the method gives a way of understanding higher-order composite defects as well.  Experimental mastery of propagating surface self-organization such as step doubling will improve over time.  Along with this improvement, we expect stochastic models like the present one to be valuable guides in achieving desired structures.

\section{Acknowledgements}
 The authors would like to thank V. Belyi, Y. Wang and S.J. Sibener for useful 
discussions. AG would also like to thank R. Bao, L. Tseng and E. Yuzbashyan for 
engaging conversations on the topic. This work was supported by the National Science Foundation via its MRSEC program under award number DMR 0213745.

\end{document}